\documentstyle[preprint,aps,eqsecnum]{revtex} 

\begin{document}

\title{
Dynamic Scaling in Epitaxial Growth
}

\author{S. Das Sarma}
\address{
Physics Department, University of Maryland, College Park, Maryland \
20742-4111
}

\maketitle

\section{Outline}
In this article I provide a brief theoretical perspective on our current
understanding of the dynamic scaling phenomena in nonequilibrium epitaxial
growth
with the emphasis on the extensively studied ultrahigh vacuum thin film
deposition growth, such as MBE growth, where the main growth front
smoothening
mechanism competing against the surface kinetic roughening induced by
inherent
shot noise fluctuations in the random deposition process is the adatom
mobility
associated with surface diffusion at the growth front.  The Introduction
sets
the tone, and in subsequent sections I expand upon the key themes arising
in the
Introduction.  In particular, the issues of universality, continuum growth
equations, atomistic growth models, and unstable growth as in smooth growth
and
mound formation are discussed in the four sections following the
Introduction.

\section{Introduction}
Epitaxy is usually thought of as an extremely smooth growth process leading
to
atomically sharp surfaces/interfaces with little ($\sim$ one atomic
monolayer)
interface roughness.  In particular, layer by layer epitaxial growth on a
singular surface takes place via two dimensional island formation where the
surface roughness oscillates between zero and one monolayer as each new
atomic
layer is formed by the deposited atoms from the incident flux with no more
than
a few ($\sim$ 1) incomplete atomic monolayers being active at the growth
front.  The reason for this extremely smooth epitaxial growth morphology is
the
high adatom mobility at the growth front enabling the incident atoms to be
``quickly'' (i.e. before the arrival of the next incident atom in its
neighborhood) incorporated at epitaxial growth sites on the surface
resulting
in the atomically flat growth morphology.  The somewhat loosely defined
``epitaxial growth temperature'' ($T_e$) is, in fact, thought to be the
temperature which is high enough for fast adatom diffusion to cause layer
by
layer (step flow) growth on a singular (vicinal) surface, but low enough
for
atomic desorption from the growth front to be unimportant.  This scenario
for
epitaxial growth is well established in the literature through {\em in
situ}
RHEED analysis (and other similar experimental techniques) of the growth
front
and through extensive computer kinetic growth simulations.

It turns out that this well-accepted paradigm for morphologically smooth
epitaxial growth is conceptually flawed from a statistical mechanical
perspective
--- the smooth epitaxial growth mode is actually unstable.  The layer by
layer
growth mode on a singular surface (or the step flow growth mode on a
vicinal
surface) is, in fact, always an initial transient which asymptotically must
lead
to kinetically rough growth \cite{1,2,3,4} at long times and large
distances. The
random fluctuations (i.e. the shot noise) inherent in the incident beam
dominate
the adatom surface diffusion at long times (i.e. after the initial layer by
layer
growth transient), and the interface roughness (as measured, for example,
by
the root mean square fluctuation in the interface width, or equivalently,
by the
number of incomplete layers dynamically active at the growth front) always
increases monotonically (until it reaches saturation at a steady state
determined by the substrate size) as a function of time after the initial
transient (during the initial transient the interface roughness oscillates
between zero and one monolayer indicating smooth epitaxial growth with an
essentially flat atomically sharp growth morphology).  It is indeed true
that
for fast enough adatom diffusion the (unstable) layer by layer growth
transient
may last for many deposited layers, and for many practical purposes this
may
suffice.  Also, for fast diffusion (i.e. high temperatures) and small enough
substrate sizes, the steady state saturated roughness may be small \cite{5}
producing a smooth surface.

In addition to unstable layer by layer and asymptotically ``stable''
kinetically rough growth modes, there has been much recent interest in a
third
surface growth scenario where mounds or pyramids dominate the growth
morphology,
and these mounds/pyramids may coarsen or steepen with time.  Such a surface
growth process via mound formation is thought to arise \cite{6} from the
so-called Ehrlich-Schwoebel (ES) barrier existing at surface step edges
which
hinders the downward adatom motion by introducing an additional step-edge
diffusion bias.  The additional step-edge diffusion bias makes it less
likely
for an adatom to attach to a down-step than to an up-step, and therefore
gives
rise to characterstic ``wedding-cake'' structures on the growing surface as
adatoms on upper terraces are prevented by the Ehrlich-Schwobel barrier
from
coming down.  The net result is a rough morphology (albeit of a particular
type
involving mounds/pyramids or similar ``wedding-cake'' structures) with the
roughness increasing monotonically in time as growth progresses.

The kinetic surface roughening phenomenon is experimentally well
established
\cite{1,2,3,4,5,6,7}.  Layer by layer epitaxial growth on a flat singular
surface
manifests itself in RHEED intensity oscillations from the growing surface
---
each oscillation corresponds to the completion of a single layer.  The
RHEED
intensity oscillation phenomenon associated with the layer by layer growth
process is so well-established that it is routinely used as a
characterization
tool in MBE growth monitoring the number of grown layers.  Experimentally
it is
known that at low temperatures growth is kinetically rough or
three-dimensional
(because the mobility of the incident atoms is low) as indicated by there
being
no RHEED intensity oscillations (the intensity drops monotonically with
time
indicating progressively rougher growth) whereas at high temperatures (when
the
adatoms are ``sufficiently'' mobile) growth occurs via the two dimensional
layer
by layer mechanism as indicated by the existence of pronounced RHEED
intensity
oscillations.  The important conceptual point is, however, the generic
observation that these pronounced RHEED intensity oscillations always decay
and
eventually disappear, indicating that the high-temperature smooth
two-dimensional layer by layer growth is an unstable transient, which
eventually
turns into kinetically rough three dimensional growth.  How long the
transient
layer by layer growth mode persists depends on the adatom mobility (which
in
turn depends exponentially on the growth temperature), and at high enough
temperatures it may be possible to grow many layers without appreciable
kinetic
roughening.  It should be noted, however, that the growth temperature
cannot be
arbitrarily high so as to avoid desorption from the growth front.  In
practice,
the growth temperature should be as high as possible (within the
restriction of
avoiding significant desorption) to obtain the best layer by layer
epitaxial
growth.  (I should mention here that most real surfaces being necessarily
slightly vicinal, RHEED intensity oscillations actually disappear at high
temperatures as the diffusion length becomes comparable to the terrace size
and
two-dimensional layer by layer growth mode gives way to the step flow
growth
mode --- in a truly high symmetry singular surface this would not happen
and the
high temperature limit then is set by the no-desorption constraint.)

One of the key recent developments in epitaxial growth phenomena has been
the
realization that the coarse-grained kinetically rough growth morphology
follows
a rather general dynamical scaling behavior.  In particular, the root mean
square fluctuation $(W)$ in the evolving surface height, $h(\mbox{\boldmath
$r$},t)$ where $h$ is the height of the interface at time $t$ for the
substrate
position $\mbox{\boldmath $r$}$, obeys the following dynamic scaling law (a
similar dynamic scaling law holds for the height-height correlation function as
well \cite{1,2,3,4}):
\begin{equation}
W(L,t) \sim L^{\alpha} f(L/\xi (t)),
\end{equation}
where
\begin{equation}
W^2 = \langle (h(\mbox{\boldmath $r$},t) - <h>)^2\rangle,
\end{equation}
is the {\em mean} squared interface width ($<h>$ being the average film
thickness, $<h> = Ft$ where $F$ is the average growth rate), $L$ is the
substrate width, and the correlation length $\xi (t)$ grows with time as
\begin{equation}
\xi (t) \sim t^{1/z},
\end{equation}
with $z$ being the dynamical exponent (which describes how correlations
spread
laterally along the substrate over time), and $\alpha$ the roughness
exponent. 
At long times, when $\xi  \sim t^{1/z} \gg L$, lateral correlations have
essentially spread over the whole substrate, and the system reaches a
saturated
steady state, indicating that the scaling function $f(x)$ must have the
long-time asymptotic form:
\begin{equation}
f(x \gg 1) \sim 1.
\end{equation}
In the short-time situation $(\xi (t) \ll L)$, however, the surface
roughness $W$
should be independent of the lateral system size $L$ because lateral
correlations
have not yet spread over the whole substrate, and therefore the short-time
asymptotic form for the scaling function $f(x)$ must be:
\begin{equation}
f(x \gg 1) \sim x^{-\alpha}.
\end{equation}
Combining the above equations one obtains the following limiting behavior
of the
dynamical surface roughness $W(L,t)$:
\begin{eqnarray}
W(L,t)&&\sim t^{\beta} \ {\rm for} \ t \ll L^z \nonumber \\
&&\sim L^{\alpha} \ {\rm for} \ t \gg L^z, 
\end{eqnarray}
where $\beta = \alpha /z$ is the growth exponent.

The physical picture underlying the dynamic scaling description is the
following.  The shot noise associated with the random deposition of the
incident
beam produces kinetic roughness in the evolving growth morphology which
leads to
monotonic increase in the interface width (roughness) $W$ as a function of
time.  In the limit of very large substrates $(L \rightarrow \infty )$
there is
no natural limiting length scale in the problem and therefore the kinetic
roughness $W \sim t^{\beta}$ increases forever.  This increase follows a
power
law, $W \sim t^{\beta}$, because there is no natural time scale in the
problem
other than that defined by the growth rate which defines the unit of time. 
Note
that the average thickness of the growing film is, by definition,
proportional
to the growth time, $<h> = Ft$, and therefore in all these scaling
relations one
could replace  $t$ by $<h>$.  For a finite substrate, a steady state is
reached
when the lateral correlations spread over the whole system $(\xi  \geq L)$
and
the interface width saturates to a steady state value, $W(L,t \gg L^z) =
W_0 \sim
L^{\beta z} \equiv L^{\alpha}$, which naturally scales as a power of the
system
size.  The key point is the existence of a lateral (dynamical) correlation
length, $\xi (t) \sim t^{1/z}$, which scales with time according to the
dynamical exponent $z$.  This is, of course, the standard scenario for
dynamic
scaling in physical phenomena.  Note that the kinetically rough surface
morphology is a self-affine fractal object which exhibits anisotropic
generic
scale invariance where the anisotropy arises from the inequivalence between
growth and substrate directions \cite{1,2,3,4}.

It should be noted that there exists no rigorous proof that a dynamic
scaling
scenario must, in general, apply to the epitaxial growth morphology.  But
there
is a great deal of empirical evidence, based on both theoretical and
simulational work as well as rapidly accumulating experimental support, for
the
applicability of dynamic scaling to epitaxial growth.  The essential idea
underlying the dynamic scaling hypothesis is that there being no
characteristic
length and time scales in the problem, dynamic scaling must hold at large
length scales and long times. I note that in all real epitaxial growth
processes there obviously are several characteristic length scales (eg. the
lattice size, the diffusion length, etc.) which determine the ``critical
regime'', i.e. the asymptotic dynamic scaling sets in at distances (and times)
much longer than the characteristic lengths (and times) naturally appearing
in
the problem.  (Calculation \cite{8} of these ``short'' distance cutoffs,
eg.
the diffusion length in MBE growth, could be quite tricky for specific
experimental situations.)  Below these ``short'' distance (time) cutoffs
dynamic
scaling does not occur, and (non-universal) transient behavior dominates. 
One
example of such ``small-scale'' transient behavior is the smooth layer by
layer
growth regime with oscillatory surface roughness occuring at length scales
smaller than $\tilde{\ell}$, where $\tilde{\ell}$ (determined by the adatom
diffusion length) is the appropriate ``coherence length'' \cite{9} for
epitaxial
growth.  (Equivalently one could define a coherence time \cite{9}.)

The usual scale invariant dynamical scaling behavior is the stable long
distance (time) asymptotic behavior of the system at large scales.  It
should
be emphasized that in many practical situations the characteristic
``short''
distance or time cutoff (eg. the coherence length or the coherence time for
layer by layer growth) may be so large that the stable asymptotic behavior
is
not of particular experimental or practical relevance, and the transient
unstable growth regime completely dominates the experimental observations. 
For
the layer by layer growth transient the ``short distance'' cutoff grows
exponentially (or faster) with temperature, for example, the diffusion
length
and therefore the coherence length for the smooth layer by layer growth
regime
increases exponentially with temperature making it possible to grow
very
smooth large area thin films without any appreciable kinetic surface
roughness.  It is to be noted, however, that the layer by layer growth
regime
is always a fnite size (time) phenomenon with kinetic surface roughening 
dominating at large scales.

\section{Universality in Kinetic Surface Roughening}
The set of exponents $\alpha , \beta ,z = \alpha /\beta$ for a particular
growth process defines a dynamic universality class which characterizes the
long wavelength asymptotic properties of kinetic surface roughening
associated
with that particular growth process.  Motivated by the central role played
by
the concept of universality in our understanding of equilibrium critical
phenomena, one naturally asks whether a few universality classes determined
by
symmetry, conservation laws, and other factors controlling growth dynamics
could describe seemingly many different surface growth processes.  In
particular, the important question is whether there are only a few possible
independent sets of exponents $(\alpha ,\beta ,z)$ describing epitaxial
growth
in different materials and under different growth conditions.  We would of
course also want to know what fundamental aspects of a growth process
uniquely
determine its universality class.  A great deal of attention has recently
been
focused \cite{1,2,3,4,5,6,7} on the issue of the possible universality
class(es) of epitaxial growth (in particular, MBE type growth).  Although
the
subject is very much in flux and no firm conclusion has yet been reached,
some
consensus has emerged in the last few years.  I summarize below the
currently
existing consensus about the four different possible dynamic universality
classes for kinetic surface roughening in epitaxial growth, and discuss the
continuum equations appropriate for each of these four universality classes
in
the next section.

\begin{enumerate}
\item {\em KPZ universality} \ If the growth process is nonconserved (i.e.
if
desorption and/or formation of surface overhang and bulk vacancies is
dynamically significant at the growth front), then the asymptotic
universality
class of the growth process is thought to belong to the Kardar-Parisi-Zhang
(KPZ) universality class \cite{10} whose exact exponents are known only in
one
substrate dimension $(d = 1 + 1)$ with $\alpha = 1/2$, $\beta = 1/3$, $z =
3/2$
in $d = 1 + 1$.  The KPZ exponents for the actual surface growth problem,
where
the substrate dimension is two $(d = 2 + 1)$, are known only approximately
(from
extensive numerical simulations) to be $\alpha \simeq 0.39$, $\beta \simeq
0.24$,
$z \simeq 1.61$.  I note that as a matter of principle all epitaxial growth
processes must be nonconserved at the largest scales of length and time
because
some desorption as well as some overhang/vacancy formation is unavoidable
in any
real growth problem.  Thus, the current consensus is that the KPZ
universality
class is, in fact, the {\em generic} growth universality class which must
apply
to all growth problems at long enough scales of lengths and times.  On the
other
hand, the KPZ exponents are almost never found in any real epitaxial
surface
growth experiments where the measured roughness $(\alpha )$ and dynamical
$(z)$
exponents tend to be much larger than the KPZ values ($\alpha \simeq 0.39$;
$z
\simeq 1.61$).  The KPZ universality class therefore has this dubious
dichotomy
of being theoretically generic for the large scale asymptotic growth
behavior
and, at the same time, being empirically inapplicable to almost all of the
currently existing experimental data on surface growth.  Part of the reason
for
this dichotomy obviously lies in the fact that in most epitaxial growth
situations, desorption from the growth front as well as vacancy/overhang
formation during growth are dynamically insignificant in the experimental
time
and length scales, and therefore the asymptotic generic universality simply
does
not manifest itself.  After all, in epitaxial growth experiments, eg. MBE
growth, one takes great care in avoiding evaporation and vacancy/overhang
formation during growth, and it is therefore perhaps not surprising that
the KPZ
exponents are never seen in epitaxial growth experiments.  Although
numerical
simulations of MBE growth allowing for defect (eg. overhangs and vacancies)
formation rather clearly demonstrate \cite{11,12,13} the expected
asymptotic
crossover to the KPZ universality, one of the most serious open issues in
this
subject has been the persistent lack \cite{14} of clear and convincing
experimental support for the existence of the generic KPZ universality in
real
epitaxial growth.
\item {\em EW universality} \ For conserved (i.e. no desorption or
vacancy/overhang formation) solid-on-solid (SOS) type growth processes the
KPZ
universality is inapplicable, and at least three different possible
conserved
SOS growth universality classes have extensively been discussed in the
literature during the last five years.  The most obvious one is the
so-called
Edwards-Wilkinson (EW) growth universality class \cite{15} which applies
whenever the growth process involves some ``downward funneling'' type
relaxation mechanism which pushes the growth front atoms down to the local
height minima \cite{16}.  Any tendency (however small) for the atoms to
preferentially attach to down-steps at terrace edges rather than to
up-steps
during growth automatically leads to the EW universality class.  (Note that
this tendency of preferential attachment to down-steps over up-steps is
precisely opposite to what happens for growth under an Ehrlich-Schwoebel
step-edge diffusion bias where atoms preferentially attach to up-steps
rather
than down-steps --- this explains why one occasionally encounters in the
literature statements to the effect that a ``negative'' Ehrlich-Schwoebel
barrier leads to EW universality.)  The EW growth universality class is
characterized by the exactly known critical exponents $\alpha = 0(1/2)$,
$\beta
= 0(1/4)$, $z = 2(2)$ in $d = 2 + 1 (1 + 1)$ dimensions.  (The EW critical
exponents for kinetic roughening are exactly known in arbitrary
dimensions.)  An
important point is that EW universality predicts ``smooth'' growth $(\alpha
=
\beta = 0)$ in the physical $2+1$-dimensional kinetic surface roughening
process, implying only logarithmic increase in the kinetic roughness with
time
and system size.  (Note that vanishing critical exponents implies
logarithmic
increase in kinetic roughening and {\em not} vanishing roughening as one
may naively assume.)  Of all the theoretically possible growth universality
classes, the EW universality class has the smoothest evolving growth
morphology
because it has the smallest possible values of the growth $(\beta )$ and
the
roughness $(\alpha )$ exponents in any dimension.  This is significant from
a
technological perspective because large area smooth thin films and surfaces
are
required for most electronic materials applications.  Understanding growth
techniques and physical mechanisms which may induce EW universality in
epitaxial
growth would be immensely valuable technologically.  It has recently been
suggested \cite{17} that surfactants induce smooth epitaxial growth by
causing
preferential attachment of deposited adatoms at down-steps and thereby
introducing the EW universality behavior $(\alpha = \beta = 0)$ in a
natural
manner.  A systematic experimental search for various techniques which
induce EW
growth could be quite useful.
\item {\em MH universality} \ In conserved SOS growth situations where
``surface diffusion'' is dynamically significant in the absence of any EW
relaxation process (i.e. no downward funneling or preferential attachment
to
down-steps), the growth process may belong to the linear surface diffusion
(the
so-called Mullins-Herring, MH) universality class \cite{18,19}.  The
critical
exponents for the MH growth universality class are exactly known
theoretically:
\ $\alpha = 1(3/2)$, $\beta = 1/4(3/8)$, $z = 4(4)$ in $d = 2 + 1 (1 + 1)$
dimensions.  Although several experimental claims in the literature report
measurements of kinetic surface roughening exponents, $\beta \approx 0.25$
and
$\alpha \approx 0.9-1$, consistent with the MH universality predictions, it
is
unclear at the present time whether the MH universality could ever be a
true
asymptotic universality class (at long times and distances) rather than
being a
short distance/time crossover phenomenon with the asymptotic universality
being
given by the EW universality defined in (2) above or the MBE universality
defined in (4) below.  The current theoretical consensus is that the MH
universality can only be a crossover and {\em not} an asymptotic
universality
class.  I note that the MH universality (sometimes also referred to as the
``linear surface diffusion'' universality) has the largest critical
exponents
$\alpha$ and $\beta$ among the four epitaxially relevant universality
classes
being discussed here, implying that MH universality produces the maximum
large-scale
kinetic surface roughness in the evolving growth morphology.  Note
that
$\alpha = 1$ for the MH universality in the surface growth problem, which
means
that the saturated steady state interface width is proportional to the
substrate
size, $W(L,t \rightarrow \infty ) \propto L$ in $d = 2+1$.  In one
dimension $(d
= 1+1)$, the roughness exponent, $\alpha = 1.5$, exceeds unity, which is a
manifestly peculiar situation implying that the large scale steady state
morphology of the growing surface is not self-affine in $d = 1+1$ because
$W(L,t \rightarrow \infty)/L$ actually diverges as $\sqrt{L}$ in the
thermodynamic limit.  This extreme pathological roughness in the evolving
growth morphology of the MH universality class effectively rules it out as
a
true asymptotic description for a physical surface growth problem even
though
it may very well dominate the short-time (non-asymptotic) kinetic surface
roughening in many experimental situations where surface diffusion plays an
important role.  I mention that the dynamic exponent, $z=4$, is very large
for the MH universality class, making the evolution of lateral correlations
extremely slow, $\xi (t) \sim t^{1/4}$, and this slow evolution of
correlations
considerably complicates studying crossover properties as well as steady
state
behavior in the MH universality problem. 
\item {\em MBE universality} \ The most relevant universality class in the
context of conserved epitaxial growth is the molecular beam epitaxy (MBE)
universality, which somewhat confusingly goes by several other names in the
literature as well: \ conserved epitaxial growth, conserved KPZ, nonlilnear
surface diffusion, and Villain-Lai-Das Sarma (VLD) or equivalently Lai-Das
Sarma-Villain (LDV) universality \cite{20,6} all describe this same
universality
I refer to as the MBE universality in this article.  This universality
class is
believed by many to be the correct large scale asymptotic description for
the
kinetic surface roughening properties of epitaxial growth morphology (which
explains my nomenclature for this universality) provided that the growth
process
is conserved SOS (no desorption and overhang/vacancy formation) and that
the
``downward funneling'' type EW relaxation processes are absent or are
dynamically unimportant.  The critical exponents for the MBE universality
class
are calculated to be \cite{20} $\alpha = 2/3(1)$, $\beta = 1/5(1/3)$, $z =
10/3(3)$ in $d = 2+1(1+1)$ dimensions.  These calculated exponents were
thought
to be exact until very recently when some questions \cite{21} have been
raised
about their exactness.  Any corrections (if there are any, the issue is by
no
means settled) to these exponent values are, however, estimated
\cite{21,22} to
be well below 1\%, and are therefore of no practical significance.  There
are
many reports in the literature of measuring critical exponents $\beta
\approx
0.2-0.3$ and $\alpha \approx 0.6-0.9$ in surface growth experiments, which
support the applicability of the MBE universality class to real epitaxial
growth
morphology.  Existing realistic MBE growth simulations based on kinetic
Monte
Carlo studies \cite{4,23} also lead to critical exponents consistent with
the
MBE universality class exponents.  It should be noted that often these
experimentally measured or numerically simulated exponents appear to be
somewhat
larger than the MBE universality exponents $(\beta = 0.2, \ \alpha =
0.67)$,
indicating that real epitaxial growth morphology may actually be exhibiting
a
crossover from the MH universality $(\beta = 0.25, \ \alpha = 1.0)$ to the
MBE
universality.  One definitive way of distinguishing between these two
universality classes would be a measurement of the surface skewness $s$
\begin{equation}
s = \langle (h-<h>)^3\rangle /\langle (h-<h>)^2\rangle^{3/2},
\end{equation}
associated with the growth morphology.  Both the EW and the MH universality
(being linear universality classes) predict a vanishing value of $s$
whereas
the MBE universality class \cite{4} predicts $s \neq 0$ in the steady state
saturated morphology (as well as in the pre-asymptotic transient regime). 
There have recently been reports \cite{24} of non-zero surface skewness
measurements in epitaxial growth morphology, again supporting the viewpoint
that the MBE universality class indeed provides the appropriate dynamic
scaling
description of epitaxial growth.  (In the context of surface skewness, I
point
out that the non-conserved KPZ universality also has a nonzero value
of $s$ in the preasymptotic transient regime in both $d = 1+1$ and $2+1$
dimensions, but in the asymptotic steady state situation the skewness
magically
vanishes, $s=0$, in $d=1+1$ dimensions by virtue of the validity of a
fluctuation-dissipation theorem which is also responsible for producing the
exact values of KPZ exponents in one substrate dimension \cite{25}.)
\end{enumerate}

\section{Continuum Theory}
Theoretical understanding of dynamic scaling in kinetic surface roughening
has
been exclusively based on stochastic partial differential equations
describing
the time evolution of the coarse-grained surface height variable
$h(\mbox{\boldmath $r$},t)$ ever since the pioneering work \cite{10} of
Kardar,
Parisi and Zhang (more than ten years ago) who showed that the well-known
(in
the context of fluid dynamics) noisy Burgers equation is the appropriate
long
wavelength asymptotic dynamical description for nonconserved generic growth
under non-SOS ballistic deposition conditions (which allow the formation of
overhangs and vacancies at the growth front).  In this section I will
briefly
discuss the continuum growth equations which define the four universality
classes described in the last section.

It is notationally more convenient to write the dynamical equations in
terms of
the height fluctuation, $h(\mbox{\boldmath $r$},t) - <h>$, which is
equivalent
to a transformation to a moving frame of reference which takes the average
interface height $<h> = Ft$ as the reference position rather than the
original
substrate $<h> = 0$.  This eliminates the trivial constant term, given by
the
average growth rate $F$, explicitly from the equation.  From now on $h$
implies
the height fluctuation (around the average) rather than the height variable
itself.  The celebrated KPZ equation \cite{10} is then the deceptively
simple-looking nonlinear stochastic equation
\begin{equation}
{\rm KPZ}: \ \frac{\partial h}{\partial t} = \nu_2 \nabla^2 h + \lambda_2
(\mbox{\boldmath $\nabla$} h)^2 + \eta ,
\end{equation}
where the spatial derivative $\mbox{\boldmath $\nabla$} \equiv
\frac{\partial}{\partial \mbox{\boldmath $r$}}$ is in the substrate plane,
and
$\eta$ is the stochastic shot noise associated with the random fluctuations
in
the incident beam.  Usually $\eta (\mbox{\boldmath $r$},t)$ is taken to be
a
Gaussian white noise of zero mean $<\eta >$ with the noise correlator
\[
\langle \eta (\mbox{\boldmath $r$}_1,t_1) \eta (\mbox{\boldmath
$r$}_2,t_2)\rangle = D \delta (\mbox{\boldmath $r$}_1 - \mbox{\boldmath
$r$}_2)
\delta (t_1 - t_2)
\]
where $D$ is the strength of the bare noise.

I do not discuss here the KPZ equation at all because, as stated in the
last
section, it is not particularly relevant for understanding epitaxial growth
morphology in experimental length and time scales.  (Several excellent
reviews
\cite{1,2,25} of the KPZ equation exist in the literature.)  Two particular
features of Eq. 4.1 should be noted: \ (a) It is the lowest order growth
equation consistent with the symmetries of the problem, and (b) the
nonlinear
$(\nabla h)^2$ term in the equation, a manifestly nonequilibrium growth
term as
it breaks the up-down symmetry in the problem, is a nonconserved term
associated
with vacancy/overhang formation (or desorption) at the growth front.

Before proceeding any further I write down below the three continuum growth
equations which correspond respectively to the three other universality
classes
introduced in the last section:
\begin{equation}
{\rm EW}: \ \frac{\partial h}{\partial t} = \nu_2 \nabla^2 h + \eta
\hspace{1in},
\end{equation}
\begin{equation}
{\rm MH}: \ \frac{\partial h}{\partial t} = -\nu_4 \nabla^4 h + \eta
\hspace{1in},
\end{equation}
\begin{equation}
{\rm MBE}: \ \frac{\partial h}{\partial t} = -\nu_4 \nabla^4 h +
\lambda_{22}
\nabla^2 (\mbox{\boldmath $\nabla$} h)^2 + \eta,
\end{equation}
Several features of these four growth equations, Eqs. 4.1-4.4,
corresponding to
the universality classes 1-4 respectively in section II, should be noted: \
the EW and the MH equations are linear and are therefore very simple; the
KPZ
(MBE) equation is simply the nonlinear version of the EW (MH) equation with
the
appropriate 2nd (4th) order nonlinearity being included in the equation;
the
MBE equation is effectively a conserved KPZ equation with a nonconserved
external noise rather than a conserved noise \cite{26} $\frac{\partial
h}{\partial t} = -\nu_4 \nabla^4 h + \lambda_{22} \nabla^2 (\mbox{\boldmath
$\nabla$} h)^2 + \eta = \nabla^2 (-\nu_4 \nabla^2 h + \lambda_{22}
(\mbox{\boldmath $\nabla$} h)^2) + \eta$; EW and KPZ equations are 2nd
order
equations while MH and MBE equations are fourth order and therefore a
growth
equation containing all of these terms will be dominated by the KPZ
nonlinearity
(if it is non-zero) or by the EW term (if $\lambda_2 = 0$) with the $\nu_4
\nabla^4 h$ and the $\lambda_{22} \nabla^2 (\mbox{\boldmath $\nabla$} h)^2$
terms being irrelevant in the renormalization group sense.  I emphasize,
however,
that even in such a situation the $\nabla^4 h$ and the $\nabla^2
(\mbox{\boldmath $\nabla$} h)^2$ terms could significantly affect crossover
properties and dominate the critical properties at ``short'' length and
time
scales.

Restricting to epitaxialy relevant conserved SOS growth (with the nonconserved
noise arising from external beam fluctuations) it is easy to see
\cite{6,20}
that the coarse-grained continuum growth equation must obey a continuity
equation consistent with the conservation of the surface current in the
problem:
\begin{equation}
\frac{\partial h}{\partial t} + \mbox{\boldmath $\nabla$} \cdot
\mbox{\boldmath
$j$} = \eta ,
\end{equation}
which follows simply from the mass and volume conservation in epitaxy
associated with the no overhang/vacancy/desorption hypothesis.  Thus the
growth
equation must have the form $\frac{\partial h}{\partial t} =
-\mbox{\boldmath
$\nabla$} \cdot \mbox{\boldmath $j$} + \eta$ where the surface current
density
$\mbox{\boldmath $j$}$ can only be of a form which is consistent with the
symmetries in the growth problem.  (It should be noted that the KPZ
equation,
Eq. 4.1, is explicitly ruled out because the KPZ nonlinearity, $(\nabla
h)^2$,
cannot be expressed as the divergence of a current.)  The only general
symmetries in the epitaxial growth problem are the translational invariance
along the growth direction (i.e. it does not matter where the substrate is
placed) and the rotational invariance in the substrate $\mbox{\boldmath
$r$}$-plane.  Taking into account these general symmetries and the
conserved
current constraint, the conserved epitaxial growth equation could be
written as
\cite{20}
\begin{equation}
\frac{\partial h}{\partial t} = \nu_2 \nabla^2 h - \nu_4 \nabla^4 h +
\lambda_{22} \nabla^2 (\mbox{\boldmath $\nabla$} h)^2 + \lambda_{13}
\mbox{\boldmath $\nabla$} (\mbox{\boldmath $\nabla$} h)^3 + \eta,
\end{equation}
keeping terms only up to the fourth order.  This is the most general
possible
low order conserved epitaxial continuum growth equation.  The conserved
universality classes discussed in the last section are all special limiting
cases of Eq. 4.6, as is obvious by comparing Eq. 4.6 with Eqs. 4.2-4.

A simple power counting analysis shows that in Eq. 4.6 the most relevant
long
wavelength term is the EW $\nabla^2h$ term, followed respectively by the
$\mbox{\boldmath $\nabla$}(\mbox{\boldmath $\nabla$} h)^3$ term, the
$\nabla^2
(\mbox{\boldmath $\nabla$} h)^2$ term, and the $\nabla^4 h$ term in order
of
most relevant to the least relevant growth processes at the largest scales.

Since asymptotic critical exponents are always determined by fluctuations
at
the largest scales, one concludes that the EW universality, if present
(i.e.
if $\nu_2 \neq 0$), is the most {\em relevant} universality in the
problem.  The other terms may still be quite important in determining the
{\em effective} exponents at shorter scales (and thus affect crossover
behavior), but the asymptotic universality class is always EW if $\nu_2
\neq 0$
in Eq. 4.6.  An additional notable point is that the most relevant fourth
order term, the $\lambda_{13} \mbox{\boldmath $\nabla$} (\mbox{\boldmath
$\nabla$} h)^3$ term, originally introduced in ref. \cite{20} through
symmetry
considerations and recently rediscovered \cite{27} in the context of
unstable
growth under an ES barrier, generates the EW $\nabla^2h$ term upon
renormalization \cite{28} and therefore also leads to EW universality
although it is formally a higher order term.  Thus even if $\nu_2 = 0$, but
$\lambda_{13} \neq 0$ in Eq. 4.6, the growth universality class is still
EW!  (I
note that the most relevant conserved term in each order, namely the
$\mbox{\boldmath $\nabla$} (\mbox{\boldmath $\nabla$} h)^{2n+1}$ term with
$2n
= 2,4,6$, etc., always generates EW universality upon renormalization ---
it
is, however, difficult to imagine physical processes which could lead to
$\nu_2
= 0$ but these nonlinear terms non-zero.)

The fourth order linear (Eq. 4.3) or nonlinear (Eq. 4.4) continuum growth
equations become applicable to conserved epitaxial growth only in the
absence of
all EW relaxation processes (i.e. when $\nu_2 = \lambda_{13} = 0$), or as
crossover effects at short distance/time scales before the EW growth terms
in
Eq. 4.6 become dynamically significant.  As mentioned in section III, there
is a
large body of experimental data \cite{1,2,4,7} on the surface kinetic roughening
in epitaxial growth reporting a roughness exponent $\alpha \sim 0.6-0.9$
and
a growth exponent $\beta \sim 0.2-0.3$, which are totally inconsistent with
the EW exponents $(\alpha = \beta = 0)$, but are consistent with the MH
growth
equation $(\alpha = 1.0$, $\beta = 0.25)$ and the MBE growth equation
$(\alpha
= 0.67$, $\beta = 0.2)$.  Empirically, therefore, it seems that epitaxial
growth belongs to the MBE universality class as defined by Eq. 4.4 (which
incorporates as a special case the MH equation, Eq. 4.3, in the special
situation
of the vanishing nonlinearity $\lambda_{22} = 0$).

Obtaining the critical exponents from the continuum growth equation is
simple
and straightforward for the linear equations (Eqs. 4.2 and 4.3) because the
linear equations can be directly solved via Fourier transformation.  The
nonlinear MBE growth equation, Eq. 4.4, is thoroughly nontrivial, however,
and
has been analyzed using the dynamical renormalization group technique
\cite{20} and the direct numerical integration technique \cite{28,29,22}. 
I
summarize below the theoretical exponent values for Eqs. 4.2-4:
\begin{equation}
{\rm EW}: \ \alpha = (3-d)/2, \ \beta = (3-d)/4, \ z = 2, \hspace{.85in}
\end{equation}
\begin{equation}
{\rm MH}: \ \alpha = (5-d)/2, \ \beta = (5-d)/4, \ z = 4, \hspace{.85in}
\end{equation}
\begin{equation}
{\rm MBE}: \ \alpha = (5-d)/3, \ \beta = (5-d)/(7+d), \ z = (7+d)/3.
\end{equation}
In Eqs. 4.7-9, $d$ is the total spatial dimension, and obviously the
physically
relevant dimension for kinetic surface roughening in epitaxial growth is
$d=2+1=3$.  The one dimensional substrate problem, $d=1+1$, is, in fact,
extensively studied theoretically (mainly because of convenience and ease
in
numerical simulations), and may have considerable potential experimental
relevance in terms of kinetic roughening of step edges \cite{14} on vicinal
surfaces.  While the above results (Eqs. 4.7 and 8) for the linear EW and
MH
universality classes are obviously exact, the MBE universality results
given in
Eq. 4.9 are obtained from a one-loop dynamical renormalization group
analysis
which is leading order in the $\epsilon$-expansion where $\epsilon = 5-d$. 
Given that the expansion parameters $\epsilon = 2(3)$ in $d=3(2)$ are
obviously
not small, one may legitimately question \cite{21} the validity of the
calculated
exponents in Eq. 4.9.  Direct numerical integration of Eq. 4.4, however,
gives
\cite{28,29,22} the same (within the numerical accuracy $\sim$ 1-2\%)
exponents
in $d=2$ and 3 dimensions as those of Eq. 4.9, providing support to the
one-loop
dynamical renormalization group result \cite{20}.  It is possible that the
exponents of Eq. 4.9 are, in fact, exact for the MBE universality class due
to
some hidden symmetry in the problem \cite{20,26}, but the issue has to be
considered open at this time \cite{21}.

Combining the current conservation condition with the general conserved SOS
growth equation (Eq. 4.6) one gets the following expression for the surface
current in epitaxial growth
\begin{equation}
\mbox{\boldmath $j$} = -(\mbox{\boldmath $\nabla$}h) \{ \nu_2 +
\lambda_{13}
(\mbox{\boldmath $\nabla$}h)^2 + \cdots \} + \mbox{\boldmath $\nabla$} \{
\nu_4 (\nabla^2h) - \lambda_{22} (\mbox{\boldmath $\nabla$}h)^2 + \cdots
\},
\end{equation}
where the first term, a nonequilibrium surface current $j_{\rm NE}$, is
proportional to the surface slope $\mbox{\boldmath $\nabla$}h$ and the
second
term can be thought of \cite{6} as the contribution from a generalized
chemical
potential $\mu$, where $\mu = \mu_0 + \mu_{\rm NE}$ with $\mu_0 \propto
\nabla^2h$ is the standard ``equilibrium'' chemical potential leading to
the
Mullins-Herring equation \cite{30} and the so-called nonequilibrium
chemical
potential $\mu_{\rm NE}$ is given by $\mu_{\rm NE} \propto \lambda_{22}
(\mbox{\boldmath $\nabla$}h)^2 + O((\mbox{\boldmath $\nabla$}h)^4) +
\cdots$. 
Thus, Eq. 4.10 can be rewritten \cite{31} as $\mbox{\boldmath $j$} =
\mbox{\boldmath $j$}_{\rm NE} + \mbox{\boldmath $\nabla$}\mu$ where
$\mbox{\boldmath $j$}_{\rm NE} \propto (\mbox{\boldmath $\nabla$}h)$.  Note
that the chemical potential contribution to the current $\mbox{\boldmath
$j$}$
does not explicitly depend on $\mbox{\boldmath $\nabla$}h$ whereas the
nonequillibrium current contribution, $\mbox{\boldmath $j$}_{\rm NE}$,
which
leads to the EW universality is proportional to $\mbox{\boldmath
$\nabla$}h$. 
This provides \cite{31} a simple and effective technique of detecting the
possible existence of the EW universality in a particular growth model.  If
there is a net inclination dependent surface current on a tilted substrate
in a
growth model, then $j_{\rm NE} \neq 0$, indicating the presence of the EW
universality in the model (this is true only if the current is downward and
therefore stabilizing --- if the inclination dependent current is upward
then
that indicates the existence of an ES barrier in the problem with the
associated
growth instability).  On the other hand, if the measured inclination
dependent
current vanishes on a tilted substrate, then obviously the EW universality
is
absent in the model.  This simple technique of measuring the inclination
dependent surface current on a tilted substrate has turned out to be a
powerful
tool for detecting the presence or absence of the EW universality in
various
growth models, and is operationally very effective in determining the EW
universality long before the measured critical exponents show a crossover
to EW
behavior \cite{2,31}.  Application of this technique of measuring
inclination
dependent surface current on tilted substrates enables one to effectively
distinguish among unstable ES barrier growth (upward current), EW
universality
(downward current), and the fourth order MBE growth universality (vanishing
inclination dependent current) in various discrete epitaxial growth models
(some
of which are discussed in the next section).

I conclude this section by mentioning that the actual values of the
coefficients $\nu_2$, $\nu_4$, $\lambda_{22}$, $\lambda_{13}$, etc.
entering
the continuum growth equations are not important in determining the
respective
asymptotic universality classes except in the trivial sense that a
particular
coefficient must be non-zero for that particular universality class to show
up.  The explicit values of these coefficients are, however, important in
determining the crossover behavior in the growth problem, for example, if
only
the (most relevant) $\nu_2$ and the (most irrelevant) $\nu_4$ terms are
non-zero in Eq. 4.6 then the growth universality class is obviously EW
since
$\nu_2 \nabla^2h$ is more relevant than $\nu_4 \nabla^4h$, but a simple
dimensional analysis shows that the EW universality can show up only for
lateral length scales larger than $(|\nu_4|/|\nu_2|)^{1/2}$ or equivalently
for
time scales larger than $(|\nu_4|/|\nu_2|)^2$, and therefore in a real
finite
size/time experiment the EW universality may simply not manifest itself,
leading
to effective MH universality exponents.  This is a situation where an
inclination dependent current measurement on a tilted substrate may lead to
the
``correct'' (i.e. EW) asymptotic universality even though the ``effective''
exponents are still in the MH crossover regime.  Finally, I point out that
the
coefficients $\nu_2$, $\lambda_{13}$, and $\lambda_{22}$ are expected to be
proportional to the external flux rate $F$ in the leading order because
these
terms are manifestly nonequilibrium contributions ($\nu_2$ and
$\lambda_{13}$
to $\mbox{\boldmath $j$}_{\rm NE}$, and $\lambda_{22}$ to $\mu_{\rm NE}$)
to the
growth process, and therefore should vanish in the absence of the external
flux.  The $\nu_4$ term, on the other hand, is the usual equilibrium
surface
diffusion contribution \cite{30} and its presence in the nonequilibrium
growth
problem is still somewhat mysterious \cite{6,19,2}.

\section{Atomistic Growth Models}
All real epitaxial growth takes place in discrete systems with atoms and
lattices whereas the coarse-grained long wavelength theory discussed in the
last section is, by definition, a continuum theory.  A question naturally
arises about whether such a coarse-graining procedure leading to a
continuum
dynamical growth equation which describes the long wavelength asymptotic
dynamic scaling properties of the growth problem is, in general, allowed
for
all discrete expitaxial growth phenomena.  The answer to this question is
not
known, and only in a few problems a rigorous connection between a discrete
growth model and its coarse-grained continuum description has been
established.  One hopes, with considerable physical justification, that
such a
coarse-graining prescription leading from a discrete problem to a continuum
description should, in fact, be possible at ``sufficiently long''
wavelengths. 
(Whether real experimental growth studies are always carried out in this
``sufficiently long'' wavelength regime or not is, of course, a totally
different issue which is extremely difficult to answer.)  Recently, some
subtle
and serious questions have been raised \cite{32,33,34} by several different
groups regarding the validity of the continuum descriptions for discrete
growth
problems, particularly in cases involving nonlinear (eg. KPZ, MBE, etc.)
growth
terms.  Although the details of these doubts vary considerably from case to
case, it is fair to say that one of the tentative conclusions emerging from
these analyses \cite{32,33,34} is that in some situations involving
nonlinear
growth processes discrete growth problems and their coarse-grained
continuum
descriptions may belong to different universality classes.  (Not
surprisingly
no such problem arises in linear growth problems, eg., EW universality.) 
There is thus sufficient reason to be critical and skeptical about blindly
applying continuum theoretical results to discrete growth problems.

The most effective technique to study atomistic epitaxial growth is via the
direct numerical simulation of the discrete growth process.  There have
been
numerous such studies using a variety of techniques and models in the
context
of dynamic scaling in epitaxy --- see, for example, refs.
\cite{4,5,11,12,13,16,18,19,22,23,34} and the review articles in refs.
\cite{1,2,3}.  In the context of epitaxy, where adatom mobility at the
growth
front is the dominant smoothening mechanism and the shot noise fluctuations
inherent in the deposition beam is the roughening mechanism, the kinetic
(or,
the stochastic) Monte Carlo simulation \cite{4,5,11,12,13,23} has been the
most realistic direct numerical technique for studying MBE growth.  In this
technique both deposition and diffusion are taken to be stochastic
processes
simulated by various random number generators (and hence the name, Monte
Carlo
simulation) with the atomic diffusion at the growth front taken to be an
Arrhenius activated hopping process with the activation energy (consistent
with the principle of detailed balance) determined by the nature of local
bonding for the hopping atom.  Depending on the local bonding of the
diffusing atom, several different activation energies (and consequently
several different diffusion rates) participate in the hopping process,
leading to the possibility of rather strong finite size and crossover
effects
in the simulation.  Thus, the full activated diffusion stochastic Monte
Carlo
simulation of epitaxial growth, while being reasonably realistic in
capturing
the evolving surface morphology, is not always the ideal method for
obtaining
the asymptotic growth exponents.

Purely nonequilibrium zero temperature growth models, with instantaneous
relaxation of the deposited adatoms according to some physically
\cite{18,19}
or mathematically \cite{16,22} motivated local rules, have been very
popular
and extremely successful in characterizing the asymptotic growth
universality
class and the dynamic scaling behavior of epitaxy.  I will discuss one such
conserved SOS nonequilibrium growth model (called the DT model by Krug
\cite{2}), introduced in ref. \cite{19} and extensively studied in refs.
\cite{4,5,35,36,37}.  Before discussing the model I mention two significant
salient features of the model which make it particularly relevant for studying
dynamic scaling in epitaxial growth: \ (1) the scaling exponents $\alpha$,
$\beta$ calculated for this simple nonequilibrium model agree {\em
quantitatively} with the corresponding exponents for the full activated
diffusion stochastic Monte Carlo simulation results in both $d=1+1$ and
$d=2+1$
dimensions \cite{4,37}; (2) due to a simple symmetry \cite{2} in the growth
rules of this model, the inclination dependent surface current vanishes
exactly
for tilted substrates, implying that this model most certainly does not
have the
most relevant EW term in its continuum description (i.e. $\nu_2,
\lambda_{13} =
0$ in this growth model) --- this particular feature makes this model
unique
among the existing nonequilibrium SOS growth models, the other two such
models
introduced in ref. \cite{16} and in ref. \cite{18} are known to belong
asymptotically to the EW universality class.

In the growth model of ref. \cite{19}, atoms are deposited randomly and
sequentially (i.e. one at a time at a randomly chosen spatial position on
the
surface) on a cubic substrate according to an average rate and within the
SOS
constraint, after deposition each atom could relax within a lateral
diffusion
length (which is most commonly taken to be just one lateral lattice
spacing)
subject to the following conditions: \ only deposited atoms with no lateral
bonds to other surface atoms are allowed to move, and the atom moves
(within
the lateral diffusion length) only to increase its local coordination
number
(i.e. lateral bonding) --- otherwise it does not move and the next atom is
deposited.  Note that these local diffusion rules allow only downward
relaxation of the adatoms (to kink and trap sites, but {\em not}
necessarily to height minima \cite{16} or to sites of maximum local bonding
\cite{18}), and in case of ties (i.e. more than one final site satisfying
the
relaxation condition) the atom moves randomly with an equal probability to
any
of the allowed final sites.  Each deposited atom is allowed to relax only
once
immediately following deposition, but more elaborate local relaxation rules
have also been considered \cite{38} with concommitant increase in the
dynamic
scaling possibilities.  Note that the particularly simple local relaxation
rules of this manifestly nonequilibrium growth model are motivated by the
actual atomistic hopping processes at the MBE growth front in the low to
intermediate growth temperatures.  The idea is that only singly bonded
atoms
(to their neighbors underneath them, an essential ingredient of conserved
SOS
growth) have appreciable surface mobility at low growth temperatures, the
hopping probability of adatoms with higher coordinations being suppressed
exponentially.  The nonequilibrium growth model of ref. \cite{19} is a
typical
example of limited mobility diffusion models which have been successful in
modeling epitaxial growth.

Extensive computer simulations (using up to 10$^{12}$ deposited atoms) of
this
nonequilibrium growth model in $d=2+1$ dimensions show \cite{37} a clear
crossover in the growth exponent $\beta$ from $\beta \approx 0.25$ at short
time
scales to $\beta \approx 0.20$ at longer times (the estimated roughness
exponent $\alpha \approx 0.6 \pm 0.1$), indicating that this model truly
belongs to the MBE growth universality, and the crossover is from the MH
universality, dominated by the $\nabla^4h$ term, to the MBE universality,
dominated by the nonlinear $\nabla^2 (\mbox{\boldmath $\nabla$}h)^2$ term,
in
Eq. 4.4.  Note that the crossover time scale to see the MBE universality is
approximately $\sim |\nu_4/\lambda_{22}|^4$ and it should show up only for
substrate sizes larger than $\sim |\nu_4/\lambda_{22}|$.  Consistent with
these
expectations, the observed crossover to MBE universality manifest itself
\cite{37} only for system sizes larger than 100 $\times$ 100 --- in smaller
systems saturation occurs before the crossover and one sees only the MH
universality ($\beta \approx 0.25$, $\alpha \approx 0.9$).  These findings
\cite{37} in this simple limited mobility nonequilibrium growth model \cite{19}
are in agreement with many experimental reports on dynamic scaling in
epitaxy
with the experimental $\beta \approx 0.2-0.3$ and the experimental $\alpha
\approx 0.6-0.9$.  I speculate that the dynamic scaling in most epitaxial
kinetic
roughening experiments is essentially in the crossover regime between the
MH
universality and the MBE universality class.  It is certainly not possible
to
rule out the asymptotic existence of an EW universality in experimental
systems
at still larger scales; all one can say is that such an eventual crossover
to
the EW universality ($\beta = \alpha = 0$) has {\em not} been reported in
the literature in the context of dynamic scaling in epitaxial growth
experiments.  Based on all of these observations, it is not unreasonable to
conclude that the simple limited mobility nonequilibrium growth model of
refs.
\cite{19,37} captures the essence of the dynamic scaling behavior in epitaxial
growth.

Before concluding this section I mention several features of the epitaxial
atomistic growth model of ref. \cite{19,37} which are of potential
importance:
\begin{enumerate}
\item Explicit calculation of the inclination dependent surface current on
tilted substrates shows a vanishing current in both $d=1+1$ and $2+1$
dimensions,
implying the non-existence of any EW growth term in the model of ref.
\cite{19,37}.
\item Very recent theoretical work \cite{39,40} using the master equation
approach find that the model of ref. \cite{19,37} actually follows the MBE
growth equation, Eq. 4.4 of this paper.  (The model of Wolf and Villain
\cite{18}, which is closely related to the model of ref. \cite{19}, is
however
found \cite{39,40} to contain the EW $\nabla^2h$ term, which is consistent
with
the earlier finding \cite{31} of a tilt dependent surface current in the
Wolf-Villain model.)
\item The model of ref. \cite{19,37} shows considerable evolving skewness
in the
growing surface (implying a breaking of the up-down symmetry under the
nonequilibrium growth condition) with the best estimate for the steady
state
skewness in the surface morphology being $s \approx -0.5$ \cite{36,37,4}. 
\item For reasons not completely understood at the present time, the model
exhibits an extremely long crossover regime \cite{19,5,4} in $d=1+1$
dimensions,
with $\beta \approx 0.37$ for upto 10$^8$ monolayers of growth on a
$L=10^4$
substrate (the value of $\alpha$ for smaller system sizes, upto $L=200$, is
$\alpha \approx 1.3$) --- thus the expected crossover to the MBE
universality
($\beta = 0.33$, $\alpha = 1.0$) is extremely slow in one substrate
dimension
where the MH exponents ($\beta = 0.375$, $\alpha = 1.5$) seem to dominate
for a
long time except that the existence of a finite skewness definitively rules
out
a linear equation as the underlying continuum description.
\item Finally, the discrete model of ref. \cite{19,37} shows an extremely
intriguing anomalous scaling \cite{35} and multiaffine scaling \cite{36}
behavior where each moment of the height-height correlation function seems
to
have its own roughness exponent, indicating the growth problem to be
similar to
the intermittency phenomenon in fluid turbulence \cite{36}.  A recent
detailed
study \cite{34} of the multiscaling phenomena \cite{36}
indicates that the 
continuum equation for the discrete nonequilibrium growth model of
ref.
\cite{19} may actually be
\begin{equation}
\frac{\partial h}{\partial t} = \left[ -\nu_4 \nabla^4h + \lambda_{22}
\nabla^2
(\mbox{\boldmath $\nabla$}h)^2 + \eta \right] + \sum_{n=2,3, \cdots}
\lambda_{2n}
\nabla^2 (\mbox{\boldmath $\nabla$}h)^{2n} ,
\end{equation}
where the terms within the square brackets comprise the MBE growth
equation,
Eq. 4.4, and the infinite order series of $\nabla^2 (\mbox{\boldmath
$\nabla$}h)^{2n}$ terms with $2n = 4,6,8,\cdots$ defines a rather complex
correction to scaling, arising from an infinite series of marginally
relevant
(irrelevant) terms in $d=1+1(2+1)$ dimensions.  The multiscaling phenomenon
seems to be transient, lasting only up to 10$^4$-10$^5$ layers of growth in
$d=2+1$ dimensions \cite{37}, but is a very long lasting transient (at least up
to 10$^9$ layers) in $d=1+1$ dimensions.  A direct experimental measurement
of
the scaling properties of the higher moments of the height-height
correlation
functions in epitaxial growth will be very interesting in this
context. 
Any expeirmental evidence for multliaffine growth, where different moments
of
the height-height correlation function scale with different exponents and
where
the step height distribution of the evolvilng morphology shows an
intermittent
stretched exponential distribution \cite{36,2,4,34,37}, would not only be
interesting from the kinetic surface roughening perspective, but
will
be of broad general interest as it will be an example of a rather
simple
stochastic nonequilibrium model exhibiting intermittency and turbulence.
\end{enumerate}

\section{Related Topics}
I will very briefly discuss two topics in this section without really doing
much justice to their importance and significance.  (Each of these topics
deserves their own theoretical chapters.)  The first topic is the role of
growth temperature in dynamic scaling of epitaxial kinetic surface
roughening. 
The second topic is the issue of dynamic scaling in ``unstable'' growth
under
ES step edge diffusion bias.  In contrast to the standard lore in the
literature, where temperature induced smooth layer by layer growth and
growth
under ES diffusion bias are thought of as phenomena quite distinct from
kinetic
surface roughening, I tend to take the view that the powerful dynamic
scaling
ideas describe all these growth scenarios with appropriate modifications
and
qualifications.
\begin{enumerate}
\item {\em Temperature effects} \ Note that the growth temperature $(T)$
plays no direct or explicit role in the dynamic scaling properties of
kinetic
surface roughening discussed so far in this article.  This is in sharp
contrast to equilibrium critical phenomena where $T$ has to be tuned close
to
the critical temperature in order to observe the critical behavior and
scaling.  Kinetic surface roughening is thus generically scale invariant
and
the dynamic scaling behavior is present at long wavelengths without any
tuning
of growth parameters.  The growth temperature does, however, play an
extremely
important role in controlling the various crossover regimes, finite size
effects, and in determining the size of the transient layer by layer smooth
growth regime.  Growth temperature sets the distance/time scales which
separate the transient and the dynamic scaling regimes.  Temperature
actually
determines the important short distance cutoff in the growth problem by
controlling \cite{8} the diffusion length $\ell$ which scales as
\begin{equation}
\ell \sim (D/F)^{\gamma}
\end{equation}
where the exponent $\gamma \sim 1/2-1/6$ depends explicitly on the 
temperature dependent minimum stable island size
and can, in principle, be
calculated \cite{8} by stochastic Monte Carlo simulations or kinetic
rate-theoretic arguments, and $D$, $F$ are the diffusion rate and the
deposition
rate respectively.  The diffusion rate $D$ depends exponentially on the
temperature via the Arrhenius activation law
\begin{equation}
D \sim k_BT e^{-E_A/k_BT},
\end{equation}
and therefore $\ell \sim \frac{T^{\gamma}e^{-\gamma E_A/k_BT}}{F^{\gamma}}$
with
the activation energy $E_A$ depending explicitly on the size of the stable
islands on the surface (i.e. how many bonds are being cut in the hopping 
process).  The important point to note is that the characteristic diffusion
length varies superexponentially $(\sim T^{\gamma}e^{-\gamma E_A/k_BT})$
with
temperature, and therefore the crossover properties of dynamic scaling are
expected to be strong functions of temperature in the appropriate
temperature
windows \cite{9,5,41,42}.

It has recently been shown \cite{9} that the short distance cutoff in MBE
growth is not the diffusion length itself, but the closely related
coherence
length $\tilde{\ell}$ which varies as a power of the diffusion length
$\ell$:
\begin{equation}
\tilde{\ell} \sim \ell^{\delta}
\end{equation}
with $\delta = 4/(5-d)$ for the MBE growth equation.  Putting $d=2(3)$ one
concludes therefore that the finite size scaling behavior of the critical
exponents $\alpha$, $\beta$, etc. in the presence of the short distance
cutoff
defined by the coherence length $\tilde{\ell}$ should show a characteristic
$L/\ell^2 (d=2+1)$ or $L/\ell^{4/3} (d=1+1)$ dependence.  Such a finite
size
scaling behavior, in particular, an approximate $L/\ell^{4/3}$ scaling of
the
effective growth exponent $\beta$ on system size and diffusion length, has
earlier been reported in $d=1+1$ \cite{5,41}.  Since the diffusion length
$\ell$
itself depends strongly on temperature, this implies a potentially strong
temperature dependence of the critical exponents arising just from
temperature
dependent finite size effects.  In particular, the dynamical exponent
$\beta$
and the roughness exponent $\alpha$ should vary between their asymptotic
values
as defined by the appropriate universality for $L/\tilde{\ell} \rightarrow
\infty$ to effectively zero as $L/\tilde{\ell} \rightarrow 0$.  For
Arrhenius
activiated full diffusion MBE model, however, the zero temperature
$(L/\tilde{\ell} = \infty )$ limit of $\beta$ is trivially 1/2,
corresponding
to pure random deposition with no relaxation because at $T=0$ the adatoms
do
not diffuse \cite{5,19}.  Such a behavior of $\beta$, varying strongly with
temperature and diffusion length, has been seen in computer simulations
\cite{5,19,41,42}, and can easily be understood as manifestations of
crossover/finite size scaling behavior in the presence of a characteristic
short-distance cutoff $\tilde{\ell}$ over which the growing surface
morphology
is essentially smooth.  In particular, for $\tilde{\ell} \geq L$ the layer
by
layer growth regime persists indefinitely \cite{9}, and $\beta , \alpha
\equiv 0$
due to finite size effect.  One can therefore think of the smooth layer by
layer
growth regime as a trivial dynamic scaling regime where finite size effects
push
down the growth and roughness exponents to zero values.  Such an idea has
been
used in the literature to define and calculate the epitaxial growth
temperature
\cite{5,42}.

The basic point of physics is that kinetic roughening (for example,
coarse-grained continuum descriptions) applies only at length scales above
$\tilde{\ell}$, and below $\tilde{\ell}$ one has smooth layer by layer
growth
transient.  Even at large scales the coefficients $\nu_2$, $\nu_4$,
$\lambda_{22}$ etc. appearing in the growth Eqs. 4.2-4 depend on the growth
temperature (and deposition rate), which could produce complex temperature
dependent crossover behavior.  Careful experimental work obtaining $\beta
(T)$ as
a function of temperature in epitaxial growth will be useful in the context
of
understanding finite size effects in dynamic scaling of epitaxy.  Such
systematic temperature dependent experimental information is lacking at the
present time.  I note that the so-called epitaxial growth temperature,
$T_e$, is
defined by the condition $\tilde{\ell} (T) = L$ in this picture, and the
superexponential temperature dependence of $\tilde{\ell}$ implies a 
weak
(sub-logarithmic) system size dependence of $T_e$ \cite{5,42}.
\item {\em Unstable growth under ES barrier} \ In nonequilibrium growth
under an ES step edge diffusion bias, where atoms preferentially attach to
ascending over descending steps, a characteristic wedding cake morpology
evolves on the surface with the formation of mounds/pyramids.  These mounds
coarsen and/or steepen with time --- in the pre-saturation regime the
typical
mound radius (i.e. the lateral mound size), $R(t)$, increases with time
according to a coarsening exponent $n$, and the mound slope, $m(t)$,
increases
with time according to a steepening exponent $\lambda$:
\begin{equation}
R(t) \sim t^n; \ m(t) \sim t^{\lambda}.
\end{equation}
A crucial point (which is not always emphasized in the literature on the
subject) is that the dynamic scaling law applies to this situation exactly
in
the same form as it does in the generically scale invariant kinetic surface
roughening problem, $W \sim L^{\alpha} f(L/\xi (t))$ with $W(t, L
\rightarrow \infty ) \sim t^{\beta}$, except here the ``surface roughness''
$W(t)$ is essentially the same as the typical mound height, $H(t)$.  Using
the
fact that the mound slope $m \sim H/R$ (i.e. mound height/mound size), one
obtains the exponent identity
\begin{equation}
\beta = \lambda + n
\end{equation}
in nonequilibrium growth under an ES barrier.  (Alternative exponent
definitions are, in principle, possible which may deviate somewhat from the
above exponent identity, but in my view the definitions of $\beta$,
$\lambda$,
and $n$ used here are the natural ones both for the experimental and the
numerical simulation purpose --- an important point to note is that
coarsening and steepening compete, as is directly implied by Eq. 6.5.)  The
dynamical exponent $z$ is essentially $1/n$ for most reasonable definitions
of
the mound size, implying that $R(t)$ for all practical purposes is the
dynamical
correlation length $\xi (t)$.

This leads to the conclusion that nonequilibrium growth on flat singular
surfaces under an ES step edge diffusion bias follows the usual two
exponent
(eg. $\beta$ and $n$) dynamical scaling behavior exactly in the same manner
as
in the kinetic roughening problem.  There has been considerable earlier
confusion in this subject based on erroneous claims in the theoretical
literature of single exponent scaling in this problem.  Note that the
evolving
mounds/pyramids etc. manifestly break generic scale invariance in this
growth
problem because obviously there are characteristic length scales (eg. the
mound
size) dominating the growth morphology.  But the dynamic scaling concept
applies here at the same level of generality as it does in the generically
scale invariant kinetic surface roughening phenomenon.

Incorporation of an ES barrier in nonequilibrium surface growth has turned
out
to be a rather formidable theoretical challenge.  There is no suitable
coarse-grained continuum growth equation which catches all aspects of the
interesting mound formation and its subsequent coarsening.  Although there
are
a number of proposed growth equations, and in some cases the exponents
$\beta$,
$n$, $\lambda$ have been calculated, there is no existing theoretical
consensus
on a continuum description for nonequillibrium surface growth under ES
barriers.  Surprisingly, numerical simulations of discrete nonequilibrium
surface growth models under ES barriers have also turned out to be quite
tricky,
with the results depending strongly on the details of how the ES barrier is
incorporated \cite{43,44,45,46,47,48,49,50,51,27}.  In fact, there is no
consensus even on the basic nature of the instability \cite{6} associated
with
surface growth under an ES bias --- in particular, much debate and
speculation
can be found in the literature on whether or not there is slope selection
(the
exponent $\lambda \rightarrow 0$ if there is slope selection) during the
coarsening of the mounds.  Experimental studies seem to indicate that there
may
or may not be slope selection depending on the systems and growth
temperatures
one uses.  I believe that a good part of the theoretical problem arises
from our
ignorance about how to incorporate the effect of ES barrier in the
coarse-grained
continuum description as well as in atomistic numerical simulations.  In
the
stochastic Monte Carlo growth simulation, an ES bias can be included
\cite{46}
either as a reflection barrier \cite{46,47,48} or as an edge barrier
\cite{46,44}.  In the reflection barrier case, an atom approaching a
descending step is reflected back onto the upper terrace and is 
probabilistically
prevented from reaching the top of the step edge.  If it is already at
the
step edge, it may, however, go down and attach itself to the descending
step
without encountering any additional barrier.  In the edge barrier case,
there is
a barrier for the atom to go down to the step edge, but there is no
barrier to
reach the top of the down-step (the ES bias only hinders it from going down
once it is already at the top of the step).
Both of these ways of incorporating ES barrier
are simplifications of the realistic atomic potential near terrace edges
\cite{52} which have been calculated in a few cases.  The correct situation
is a
complicated combination of edge and reflection barriers, neither by itself
is
adequate.  Comparing with the realistic surface potential contours
\cite{52},
however, it is quite obvious that the edge barrier model is a much better
representation of the actual ES bias than the reflection barrier model. 
For
reasons not very clear to me, most of the numerical growth simulations
\cite{47,48} utilize the reflection barrier approach (and make additional
assumptions regarding the nature of surface relaxation to obtain slope
selection).  I myself believe that the edge barrier model is better suited
to
study surface growth under an ES bias than the reflection barrier model. 
Below
I discuss our recent results \cite{43} for nonequilibrium surface growth
under
an edge surface diffusion bias.

I conclude by providing the results of some very recent numerical
calculations
\cite{43} of the exponents $\beta$, $\lambda$, and $n$ in the
nonequilibrium
growth under an ES barrier in the $1+1$ and $2+1$ dimensional SOS growth
model. 
Our finding \cite{43} is that the growth exponent $\beta$ is 1/2 for this
problem (both in $d=2+1$ and $1+1$ dimensions), independent of the strength
of
the ES edge barrier.  It has earlier been pointed out \cite{38,44,45,46}
that for
a strong ES barrier, the growth exponent tends toward 1/2, but our new
finding is
that $\beta = 1/2$ always under an edge ES barrier, except the crossover
time to
observe this  asymptotic $\beta (=0.5)$ is extremely long for weak ES
barriers
which is why the existing ES barrier simulations have not always
unambiguously
observed this asymptotic regime.  Note that $\beta = 1/2$ here does not
imply
uncorrelated random Poisson growth because $n = z^{-1} \approx 0.2-0.1$. 
The
other question of considerable interest \cite{47,48} to both theory and
experiment is the issue of slope selection in nonequilibrium growth under
an ES
barrier, i.e. whether $\lambda = 0$ (i.e. $\beta = n$) asymptotically in
this
growth problem, and some magic slope is selected by the mounds which does
not
change with time and remains fixed.  We find that within an SOS model,
there is
no slope selection $\lambda \neq 0$ \cite{43} in the edge barrier model,
and any
experimental observation of slope selection must derive either from the
physics
of crystallographic orientations \cite{31} or from some other processes
\cite{47,48} which are not essential ingredients of the ES barrier physics
in
the SOS model.  My own belief is that, although slope selection may happen
at
extremely long times when the typical mound slope $m(t)$ is very large, it
is not
a generic phenomenon at small slopes.  We find $\lambda \approx 0.2-0.4$,
$n
\approx 0.2-0.1$, and $\beta \approx 0.5$ in all our edge barrier
simulations
without encountering any slope selection in either $1+1$ or $2+1$
dimensions. 
\end{enumerate}

\section{Concluding Remarks}
In this article I have provided a theoretical review from my own
perspective of
our current understanding of dynamic scaling in nonequilibrium epitaxial
growth on flat singular substrates in the high symmetry directions.  The
underlying theme has been the successful application of the dynamic scaling
hypothesis, defined by two independent critical exponents, to a wide
variety of
epitaxial growth phenomena.  I have tried to emphasize how local discrete
rules for deposition and relaxation under nonequilibrium conditions lead at
large scales to complex growth morphologies which obey dynamic scaling
according to coarse-grained continuum equations.  The subject is vast and
necessarily my perspective is based on the work I am most familiar with. 
The
list of references is by no means comprehensive.  The cited review articles
\cite{1,2,3,7,25} as well as refs. \cite{4,5} provide more references to
the literature.

\section{Acknowledgement}
I am most grateful to the United States Office of Naval Research for their
steadfast support of my research in this area.  In particular, I wish to
take
this opportunity to thank Dr. Larry Cooper (ONR) for sponsoring my research
on
MBE growth.

\end{document}